\documentclass[aps,preprint]{revtex4}%
\usepackage{amssymb}
\usepackage{amsfonts}
\usepackage{amsmath}
\usepackage{graphicx}%
\providecommand{\U}[1]{\protect\rule{.1in}{.1in}}

\begin{document}
\preprint{ }
\title[Reselience of SSM]{Resilience of the Spectral Standard Model}
\author{Ali H. Chamseddine$^{1,3}$and Alain Connes$^{2,3,4}$}
\affiliation{$^{1}$Physics Department, American University of Beirut, Lebanon}
\affiliation{$^{2}$College de France, 3 rue Ulm, F75005, Paris, France}
\affiliation{$^{3}$I.H.E.S. F-91440 Bures-sur-Yvette, France}
\affiliation{$^{4}$Department of Mathematics, The Ohio State University, Columbus OH 43210 USA}

\begin{abstract}
We show that the inconsistency between the spectral Standard Model and the
experimental value of the Higgs mass is resolved by the presence of a real
scalar field strongly coupled to the Higgs field. This scalar field was
already present in the spectral model and we wrongly neglected it in our
previous computations. It was shown recently by several authors, independently
of the spectral approach, that such a strongly coupled scalar field
stabilizes the Standard Model up to unification scale in spite of the low value
of the Higgs mass. In this letter we show that  the noncommutative
neutral singlet modifies substantially the RG analysis, invalidates our previous prediction of Higgs mass in the range 160--180 Gev,  and restores the consistency of the noncommutative geometric
model with the low Higgs mass.

\end{abstract}
\volumeyear{year}
\volumenumber{number}
\issuenumber{number}
\eid{identifier}
\maketitle





\section{Introduction}

The noncommutative geometry framework representing space-time as a product of
a continuous four-dimensional manifold by a finite discrete space has provided
a geometric explanation \cite{Why} for the intricate complexity of the
standard model of particle physics coupled to gravity, including all its fine
features such as the full Higgs sector whose geometric origin then becomes
obvious. Despite the many successes of this noncommutative model, one of its
predictions coming from using the spectral action at unification scale
concerning the Higgs mass, turns out to be contradicted by recent experimental
results. At some time, it was indeed claimed that the Higgs mass must lie in
the range of $160$ to $180$ Gev. Now that experimental evidence shows that the
Higgs mass is of the order of $125$ GeV, there are two difficulties that the
spectral model needs to resolve in order to survive as a model of gravity
unified with the Standard model in its actual experimentally validated form. The
above discrepancy between $125$ and $160-180$ Gev could be brushed away due to the
enormous extrapolation involved in running down from unification scale of
$10^{17}$ Gev to the electroweak scale, arguing that the order of magnitude of
the mass is correct while getting the precise value was too much to ask. But a
second difficulty is much more problematic: since such a low mass of the Higgs
creates an instability in the Higgs potential (where the quartic coupling of
the Higgs becomes negative at high energy) thus ruling out the
\textquotedblleft big desert" hypothesis which we were using, and invalidating
the positivity of the coupling at unification which is an essential prediction
of the spectral action.

The main result of this paper is that the full spectral action as computed in
our previous paper \cite{framework}, published in 2010, in fact already
contains the solution to this second difficulty and at the same time shows the
compatibility of the spectral model with the above low experimental Higgs
mass. The fact is that in our previous prediction of the Higgs mass, we
assumed, incorrectly, that the additional real singlet field $\sigma$
responsible for the neutrino Majorana masses \cite{framework}
would not interfere much with the
Higgs mass prediction and could be ignored. We became aware of the non-trivial
role played by such a scalar field when we realized that it has the same
structure of couplings as in the papers by various authors \cite{Elias1},
\cite{Elias2}, \cite{Darkon}, \cite{lebedev}, \cite{Bezr} who proposed to solve the above
instability problem of the Standard Model precisely by adding a new scalar
field strongly coupled to the Higgs. Their result shows that, adding this new scalar
field, everything would
be fine, with a Higgs of 125 Gev, for SM at high energies. It turns out that
by miracle their proposed couplings are exactly the same as the ones which
were delivered before by the spectral action in \cite{framework}.

\section{Higgs-Singlet scalar potential}

The spectral action for Standard Model was calculated for a cutoff function
starting with the Dirac operator of the corresponding noncommutative space,
using heat kernel methods. The part involving the Higgs and singlet fields is
given by \cite{framework}
\begin{align}
&  -\frac{2}{\pi^{2}}f_{2}\Lambda^{2}%
{\displaystyle\int}
d^{4}x\sqrt{g}\left(  \frac{1}{2}a\overline{H}H+\frac{1}{4}c\,\sigma
^{2}\right)  \nonumber\\
&  +\frac{1}{2\pi^{2}}f_{0}%
{\displaystyle\int}
d^{4}x\sqrt{g}\left(  b\left(  \overline{H}H\right)  ^{2}+a\left\vert
\nabla_{\mu}H_{a}\right\vert ^{2}+2e\overline{H}H\,\sigma^{2}+\frac{1}%
{2}d\,\sigma^{4}+\frac{1}{2}c\left(  \partial_{\mu}\sigma\right)  ^{2}\right)
\end{align}
where $H$ is the Higgs doublet and $\sigma$ the real scalar singlet associated
with the Majorana mass of the right-handed neutrino. The coefficients $f_{0}$
and $f_{2}$ are related to the spectral function by the relations
$f_{0}=f\left(  0\right)  $ and $f_{2}=%
{\displaystyle\int_{0}^{\infty}}
f\left(  u\right)  du.$ The coefficients $a,b,c,d$ and $e$ are related to the
fermionic Yukawa couplings and Majorana mass matrix. To simplify the analysis,
we shall work in the rough approximation where the Yukawa couplings of the top
quark $k^{u}$ and the neutrino (both Dirac $k^{\nu}$ and Majorana $k^{\nu_{R}%
}$ ) are dominant and in addition we introduce the dimensionless constant $n$
defined by the relation
\begin{equation}
k^{\nu}=\sqrt{n}\,k^{u}%
\end{equation}
In this approximation we have
\begin{align}
a &  =\left\vert k^{u}\right\vert ^{2}(n+3)\\
b &  =\left\vert k^{u}\right\vert ^{4}(n^{2}+3)\\
c &  =\left\vert k^{\nu_{R}}\right\vert ^{2}\\
d &  =\left\vert k^{\nu_{R}}\right\vert ^{4}\\
e &  =n\left\vert k^{\nu_{R}}\right\vert ^{2}\left\vert k^{u}\right\vert ^{2}%
\end{align}
It should be clear that there is some remaining flexibility especially in the
Majorana matrix $k^{\nu_{R}}$ for the general treatment involving in full the
three families of fermions. We work in the unitary gauge where three scalars
of the complex Higgs doublet $H$ are gauged away allowing us to set the field
$H=\left(
\begin{array}
[c]{c}%
0\\
h
\end{array}
\right)  $ where $h$ is real. It is more transparent to work with the rescaled
fields
\begin{equation}
\overline{h}=\left\vert k^{u}\right\vert h,\qquad\overline{\sigma}=\left\vert
k^{\nu_{R}}\right\vert \sigma
\end{equation}
so that the action for scalar fields reduces to the form
\begin{align}
&  -\frac{2}{\pi^{2}}f_{2}\Lambda^{2}%
{\displaystyle\int}
d^{4}x\sqrt{g}\left(  \frac{1}{2}\left(  n+3\right)  \overline{h}^{2}+\frac
{1}{4}\overline{\sigma}^{2}\right)  \nonumber\\
&  +\frac{1}{2\pi^{2}}f_{0}%
{\displaystyle\int}
d^{4}x\sqrt{g}\left(  \left(  n^{2}+3\right)  \overline{h}^{4}+\left(
n+3\right)  \left(  \partial_{\mu}\overline{h}\right)  ^{2}+2n\overline{h}%
^{2}\,\overline{\sigma}^{2}+\frac{1}{2}\,\overline{\sigma}^{4}+\frac{1}%
{2}\left(  \partial_{\mu}\overline{\sigma}\right)  ^{2}\right)
\end{align}
The gauge fields kinetic terms are normalized by setting the coefficient
$f_{0}$ to be $\frac{f_{0}}{2\pi^{2}}=\frac{1}{4g^{2}}.$ The scalar fields
kinetic terms are then normalized by rescaling
\begin{align}
\overline{h} &  \rightarrow\overline{h}\sqrt{\frac{2}{n+3}}g\\
\overline{\sigma} &  \rightarrow2\overline{\sigma}g
\end{align}
so that the Higgs-singlet potential reduces to
\begin{equation}
V=\frac{1}{4}\left(  \lambda_{h}\overline{h}^{4}+2\lambda_{h\sigma}%
\overline{h}^{2}\overline{\sigma}^{2}+\lambda_{\sigma}\overline{\sigma}%
^{4}\right)  -\frac{2g^{2}}{\pi^{2}}f_{2}\Lambda^{2}\left(  \overline{h}%
^{2}+\ \overline{\sigma}^{2}\right)
\end{equation}
where
\begin{align}
\lambda_{h} &  =\frac{n^{2}+3}{\left(  n+3\right)  ^{2}}\left(  4g^{2}\right)
\label{lambdah}\\
\lambda_{h\sigma} &  =\frac{2n}{n+3}\left(  4g^{2}\right)
\label{lambdahsigma}\\
\lambda_{\sigma} &  =2\left(  4g^{2}\right)  \label{lambdasigma}%
\end{align}
The singlet has a strong coupling $\lambda_{\sigma}=$ $8g^{2}.$ The coupling
$\lambda_{h\sigma}$ vanishes for $n=0$ and increases to $8g^{2}$ as
$n\rightarrow\infty$. The coupling $\lambda_{h}$ decreases from $\frac{4}%
{3}g^{2}$ to $g^{2}$ for $n$ varying from $0$ to $1$ and increases again to
$4g^{2}$ for $n\rightarrow\infty$.

\section{Running the initial conditions for the model with cutoff function}

Writing the RG equations for the top quark, neutrino, Higgs and singlet
quartic couplings we have \cite{RGEsing}
\begin{align}
\frac{dk^{t}}{dt}  &  =\frac{k^{t}}{32\pi^{2}}\left(  -\left(  \frac{17}%
{6}g_{1}^{2}+\frac{9}{2}g_{2}^{2}+16g_{3}^{2}\right)  +9\left(  k^{t}\right)
^{2}+2\left(  k^{\nu}\right)  ^{2}\right) \\
\frac{dk^{\nu}}{dt}  &  =\frac{k^{\nu}}{32\pi^{2}}\left(  -\left(  \frac{3}%
{2}g_{1}^{2}+\frac{9}{2}g_{2}^{2}\right)  +6\left(  k^{t}\right)
^{2}+5\left(  k^{\nu}\right)  ^{2}\right) \\
\frac{d\lambda_{h}}{dt}  &  =\frac{1}{16\pi^{2}}\left(  \left(  12\left(
k^{t}\right)  ^{2}+4\left(  k^{\nu}\right)  ^{2}-\left(  3g_{1}^{2}+9g_{2}%
^{2}\right)  \right)  \lambda_{h}\right. \nonumber\\
&  \qquad\left.  +2\left(  12\lambda_{h}^{2}+\lambda_{h\sigma}^{2}+\frac
{3}{16}\left(  g_{1}^{4}+2g_{1}^{2}g_{2}^{2}+3g_{2}^{4}\right)  -3\left(
k^{t}\right)  ^{4}-\left(  k^{\nu}\right)  ^{4}\right)  \right) \\
\frac{d\lambda_{h\sigma}}{dt}  &  =\frac{\lambda_{h\sigma}}{16\pi^{2}}\left(
\frac{1}{2}\left(  12\left(  k^{t}\right)  ^{2}+4\left(  k^{\nu}\right)
^{2}-3g_{1}^{2}-9g_{2}^{2}\right)  +4\left(  3\lambda_{h}+\frac{3}{2}%
\lambda_{\sigma}+2\lambda_{h\sigma}\right)  \right) \\
\frac{d\lambda_{\sigma}}{dt}  &  =\frac{1}{16\pi^{2}}\left(  8\lambda
_{h\sigma}^{2}+18\lambda_{\sigma}^{2}\right)
\end{align}
To run these equations, we have to run first the gauge couplings $\alpha
_{1},\alpha_{2}$ and $\alpha_{3}$
\[
{\beta_{g_{i}}=(4\pi)^{-2}\,b_{i}\,g_{i}^{3},\ \ \ {\hbox{ with }}%
\ \ b=(\frac{41}{6},-\frac{19}{6},-7),}%
\]
so that the inverse couplings are linear functions of $u=\log\,\frac{\Lambda}{M_{Z}}$ as follows
\begin{align*}
\alpha_{1}^{-1}(\Lambda)  &  =\,\alpha_{1}^{-1}(M_{Z})-\frac{41}{12\pi}%
\,\log\,\frac{\Lambda}{M_{Z}}\\
\alpha_{2}^{-1}(\Lambda)  &  =\,\alpha_{2}^{-1}(M_{Z})+\frac{19}{12\pi}%
\,\log\,\frac{\Lambda}{M_{Z}}\\
\alpha_{3}^{-1}(\Lambda)  &  =\,\alpha_{3}^{-1}(M_{Z})+\frac{42}{12\pi}%
\,\log\,\frac{\Lambda}{M_{Z}}%
\end{align*}
 where $M_{Z}$ is the mass of the $Z^{0}$ vector boson.

It is known that the predicted unification of the coupling constants does not
hold exactly. {The non meeting of the three gauge couplings, to within few
percents, is an indication that there is some missing ingredient in our
considerations, which may be related to the use of the cutoff function in the
asymptotic heat kernel expansion of the spectral action \cite{ali}. There are
some indications that a slight deviation from the cutoff function would alter
the relation }%
\begin{equation}
\frac{5}{3}g_{1}^{2}=g_{2}^{2}=g_{3}^{2}%
\end{equation}
which gets modified to depend on the slope of the derivative of the spectral
function. We will not pursue this issue in what follows, and assume that there
is some uncertainty in the actual value of the unification scale, so that the variable $u_{\rm unif}=\log\,\frac{\Lambda_{\rm unif}}{M_{Z}}$
will be tested in the range $(25,35)$ corresponding to unification scale $s$ in the range $m_Z e^{25}\sim 6.55245\times 10^{12}GeV$ to $m_Z e^{35}\sim 1.44327\times 10^{17}GeV$.

The RG equations are run down with initial conditions at unification scale for the top quark coupling
$k^{t},$ the neutrino coupling $k^{\nu}=\sqrt{n}\, k^{t}$,  the gauge
couplings, and the quartic scalar couplings as prescribed by equations \eqref{lambdah}, \eqref{lambdahsigma},
\eqref{lambdasigma}.

The effective top quark Yukawa coupling is given at unification scale by
(compare with \cite{mc2})
\begin{equation}
k^{t}(u_{\rm unif})=\sqrt{\frac{4}{n+3}}g
\end{equation}
so the top quark  mass at low scale is
\begin{equation}
m_{t}\left(  M_{Z}\right)  =k^{t}\left( 0\right)  \frac{246}{\sqrt{2}%
}=173.94\,k^{t}\left(  0\right)  \text{ Gev}%
\end{equation}
which should be compared to the experimental value
which is reached for $k^{t}\left(  M_{Z}\right)
=0.99.$
\begin{figure}
\begin{center}
\includegraphics[scale=1]{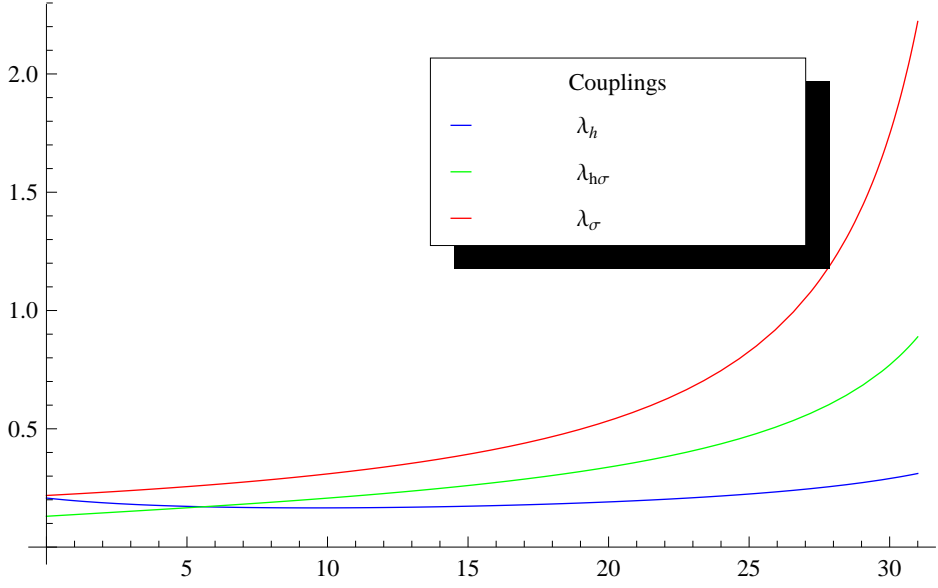}
\caption{Running of the three parameters
$\lambda
_{h},$ $\lambda_{h\sigma}$ and $\lambda_{\sigma}$}\label{run}
\end{center}
\end{figure}

\begin{figure}
\begin{center}
\includegraphics[scale=1]{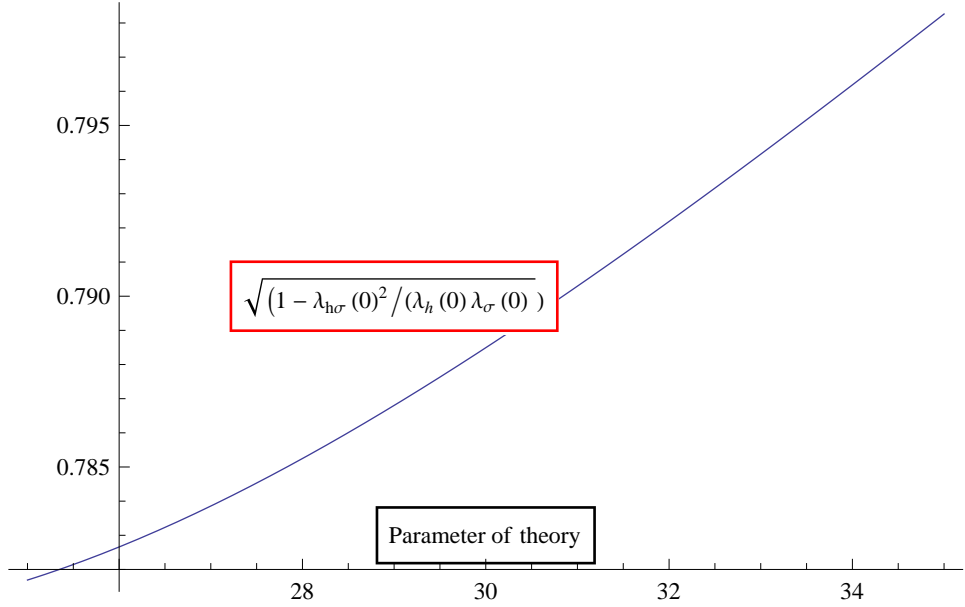}
\caption{Correction factor at scale $u=0$ as a function of the parameter of the theory}\label{run3}
\end{center}
\end{figure}

 We know that the quartic couplings $\lambda
_{h},$ $\lambda_{h\sigma}$ and $\lambda_{\sigma}$ all run, and since they are dimensionless we can trust their running
and obtain their value at low scale ($u=0$) from the initial condition at unification scale
$u_{\rm unif}$. On the other hand, the mass terms
give quadratic divergences, and their running cannot be trusted. We write the potential in the form
\begin{equation}\label{pot}
 V=  -\frac{1}{2}\left( \mu ^2\overline{h}^2 +\nu ^2 \overline{\sigma} ^2\right)+\frac{1}{4} \left(\overline{h}^4 \lambda _h+2 \overline{h}^2 \overline{\sigma} ^2 \lambda _{\text{h$\sigma $}}+\overline{\sigma} ^4 \lambda _{\sigma }\right)
\end{equation}
The minimum is obtained for
$
\left\langle \overline{h}^{2}\right\rangle    =\overline{v}^{2}$, $
\left\langle \overline{\sigma}^{2}\right\rangle    =\overline{w}^{2}
$
where
\begin{equation}\label{min}
    -\mu ^2+\bar{v}^2 \lambda _h+\bar{w}^2 \lambda _{\text{h$\sigma $}}=0, \ \
    -\nu ^2+\bar{v}^2 \lambda _{\text{h$\sigma $}}+\bar{w}^2 \lambda _{\sigma }=0.
\end{equation}
We determine the mass terms $\mu^2$ and $\nu^2$ at low scale using this equation
  so that
$\overline{v}^{2}$ is of order $10^{2}$ Gev and $\overline{w}^{2}$ is of order
$10^{11}$ Gev. We will not worry about this fine tuning, which is related to
the problem of quadratic divergencies, but note that the dilaton field \cite{scale} which
replaces the scale $\Lambda$ could be used to technically solve this problem
as it connects the different subalgebras of the discrete space. We use the
approximation $\overline{v}^{2}\ll\overline{w}^{2}$ which is also made use of  to get the see-saw mechanism. To find the masses, we
expand the scalar fields around the vev's
\begin{equation}
\overline{h}=\overline{v}+\overline{\phi},\qquad\overline{\sigma}=\overline
{w}+\overline{\tau}%
\end{equation}
and use \eqref{min} to expand the potential \eqref{pot} up to terms of third order in $ \bar{\phi }$, $\bar{\tau }$ as
\begin{equation}\label{potexp}
    V\sim \left(-\frac{1}{4} \bar{v}^4 \lambda _h-\frac{1}{2} \bar{v}^2 \bar{w}^2 \lambda _{\text{h$\sigma $}}-\frac{1}{4} \bar{w}^4 \lambda _{\sigma }\right)+
    \bar{v}^2 \bar{\phi }^2 \lambda _h+2 \bar{v} \bar{w} \bar{\tau } \bar{\phi } \lambda _{\text{h$\sigma $}}+\bar{w}^2 \bar{\tau }^2 \lambda _{\sigma }
\end{equation}
This expansion gives the mass terms for the fields $\overline{\phi}$ and
$\overline{\tau}$ in the form of the mass matrix
\begin{equation}
\frac{1}{2}\left(
\begin{array}
[c]{cc}%
\overline{\phi} & \overline{\tau}%
\end{array}
\right)  M^{2}\left(
\begin{array}
[c]{c}%
\overline{\phi}\\
\overline{\tau}%
\end{array}
\right)
\end{equation}
where
\begin{equation}
M^{2}=2\left(
\begin{array}
[c]{cc}%
\lambda_{h}\overline{v}^{2} & \lambda_{h\sigma}\overline{v}\overline{w}\\
\lambda_{h\sigma}\overline{v}\overline{w} & \lambda_{\sigma}\overline{w}^{2}%
\end{array}
\right)
\end{equation}
The eigenvalues of the mass matrix are
\begin{equation}
m_{\pm}^{2}=\lambda_{h}\overline{v}^{2}+\lambda_{\sigma}\overline{w}^{2}%
\pm\sqrt{\left(  \lambda_{h}\overline{v}^{2}-\lambda_{\sigma}\overline{w}%
^{2}\right)  ^{2}+4\lambda_{h\sigma}^{2}\overline{v}^{2}\overline{w}^{2}}%
\end{equation}
With the approximation $\overline{v}^{2}\ll\overline{w}^{2}$ we have
\begin{align}
m_{+}^{2} &  \simeq2\lambda_{\sigma}\overline{w}^{2}+2\frac{\lambda_{h\sigma
}^{2}}{\lambda_{\sigma}}\overline{v}^{2}\\
m_{-}^{2} &  \simeq2\lambda_{h}\overline{v}^{2}\left(  1-\frac{\lambda
_{h\sigma}^{2}}{\lambda_{h}\lambda_{\sigma}}\right)
\end{align}
Thus the Higgs mass is reduced by a factor of $\sqrt{\left(
1-\frac{\lambda_{h\sigma}^{2}}{\lambda_{h}\lambda_{\sigma}}\right)  }.$
This factor will be of the order $.78$ (at low scale) as shown in Figure \ref{run3}.

The condition to have a stable Higgs mass at $125$ Gev is that the determinant
of the mass matrix is positive
\begin{equation}\label{inequ}
\lambda_{h\sigma}^{2}<\lambda_{h}\lambda_{\sigma}%
\end{equation}
and we  check numerically that it holds at low scale as can be seen in Figure \ref{run3}. The physical states are mixtures of the
fields $h$ and $\sigma$ but with very small mixing of order of $\frac
{\overline{v}}{\overline{w}}=\mathrm{O}\left(  10^{-9}\right)  .$

\begin{figure}
\begin{center}
\includegraphics[scale=1]{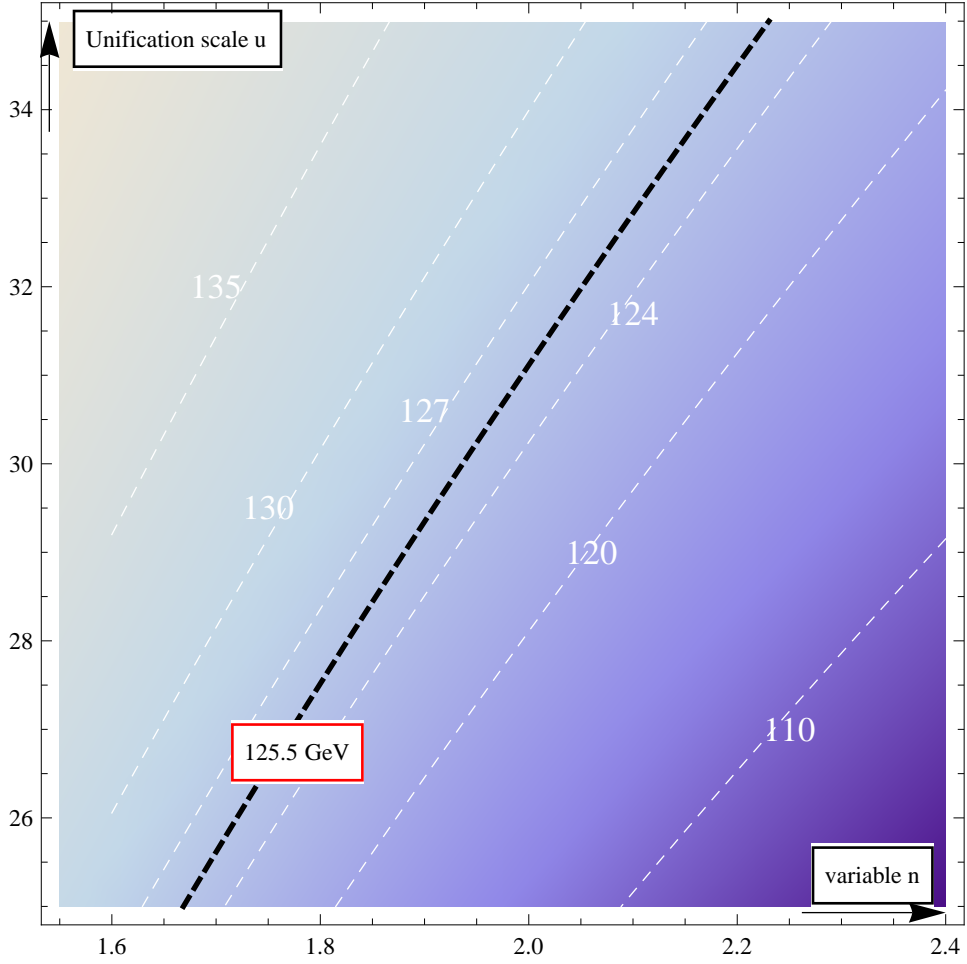}
\caption{Higgs mass as a function of $n$ and of the unification scale $u\in (25,35)$, the thick doted line is where $m_H=125.5$ Gev. The thin dotted lines correspond to constant values of $m_H$ as indicated.}\label{run2}
\end{center}
\end{figure}

Thus, we vary the parameter $n$ and the  unification scale $u_{\rm unif}$. The physical masses of the top and Higgs fields are
then determined from the values of the couplings at low energies (for $u=0$) by the formulas%

\begin{align}
m_{t}\left(  0\right)   &  =k^{t}\left(  0\right)  \frac{246}{\sqrt{2}}\\
m_{h}\left(  0\right)   &  =246\sqrt{2\lambda_{h}\left(  0\right)  \left(
1-\frac{\lambda_{h\sigma}^{2}(0)}{\lambda_{h}(0)\lambda_{\sigma}(0)}\right)  }%
\end{align}
Numerical studies of this system of one loop RG equations in the parameter space $(n,u)$ reveal that a Higgs mass
 of around 125.5 Gev is reached on an almost straight curve as shown as the thick dotted line
in Figure \ref{run2}. This shows that one can find a suitable value $n(u)$ of the free parameter $n$ for any unification scale $u$ in the range $u\in (25,35)$
(which corresponds to the interval $6.5\times 10^{12}-1.4\times 10^{17}GeV$) such that the Higgs mass has the correct experimental value. We thus obtain a one parameter family, parameterized by $u\in (25,35)$  of consistent theories. One can check numerically that for all of them  the three couplings $\lambda_{h}$, $\lambda_{h\sigma}$, $\lambda_{\sigma}$ remain positive in the running from the unification scale to the low scale and that moreover the inequality \eqref{inequ} holds at low scale (Figure \ref{run3}).

 The numerical study also shows that the  top quark mass obtained is  a few percents lower than the experimental value. It is known, however, for the two loop equations of the standard model,
without the singlet, that at the two loop level the quartic
Higgs coupling gets a very small correction, while the top quark gets a
sizable correction which is 16\% of the one loop QCD\ corrections, which are
not negligible, and which pushes the top quark mass to be higher at low
energies \cite{Elias2}, \cite{Zoller}. We thus expect, once the two loop
RG\ equations are worked out in the presence of a singlet, to improve the agreement of our theory with the
experimental values for both the top quark mass and the Higgs mass.

\section{Conclusions}

Now that the Brout-Englert-Higgs \cite{HiggsBoson} field has been discovered experimentally, it  begs for a conceptual explanation of the Lagrangian of the Standard Model  coupled to gravity, which would unify its juxtaposed fragmented pieces.
The spectral model provides such an explanation based on two ingredients:
\begin{itemize}
  \item An extension of the geometric paradigm treating the continuum and the discrete on the same footing.
  \item A principle of utmost simplicity, the spectral action principle, asserting that the action functional is a spectral function of the line element.
 \end{itemize}
The model of space-time given by the product of a continuous four dimensional
manifold times a noncommutative discrete space has many advantages.
  It allowed us to obtain  the Standard Model with the correct representation for the
$16$ fermions, the correct gauge fields  and a Higgs doublet associated with the discrete dimension,
thus providing a geometric origin for the Higgs field as a gauge field associated to the finite space.
While in the first stage of development of the model, the finite space was introduced by hand, we later classified \cite{Why} these finite spaces. We found that there are severe
 restrictions on its form, singling out, almost uniquely,
the symmetries of the standard model \cite{Why}.  A classification of finite
spaces  consistent with the axioms of noncommutative geometry and which
avoids the fermion doubling problem in Euclidean spaces showed that the
dimension of the Hilbert space of Fermions is the square of an even integer. Among very
few choices for the lowest dimensional case we obtained the algebra
$\mathcal{A}=M_{2}(\mathbb{H})\oplus M_{4}(\mathbb{C})$ where $\mathbb{H}$ is
the skew field of quaternions. This determines the number of fermions to be
$4^{2}=16$, and thus confirms the existence of the
right-handed neutrino. In addition to the Higgs field, there exists a neutral
singlet field, whose vev gives Majorana mass to the right-handed neutrino. The
existence of the singlet is responsible for the breakdown of the symmetry of
the discrete space from $\mathbb{H\oplus H}\oplus M_{4}\left(  \mathbb{C}%
\right)  $ to $\mathbb{C\oplus H}\oplus M_{3}\left(  \mathbb{C}\right)  $ and thus
plays a central role. The
noncommutative approach predicts all the fermionic and bosonic spectrum of the
standard model, and the correct representations. One can also take as a
prediction that there are no other particles to be discovered, except for the
three scalar fields: the Higgs field, the singlet field and the dilaton field
\cite{scale}. The dynamics of the fields are governed by the interactions
obtained from the spectral action principle, which is based on using a
function of the Dirac operator defining the metric of the noncommutative space. Despite the
many successes of the noncommutative model, one of the predictions concerning
the Higgs mass turned out to be problematic. At some time, it was claimed that
the Higgs mass must lie in the range of $160$ to $180$ Gev. This claim was
made under the assumption that the singlet field was integrated out, and
replaced by its vev. This assumption turns out to be simplistic. In fact the
singlet responsible for the right-handed neutrino mass gets mixed in a
non-trivial way with the Higgs field. The potential was derived and given in
full detail in our earlier work \cite{framework}. Recently and in more than
one work \cite{Elias1}, \cite{Elias2}, \cite{Darkon}, \cite{lebedev}, it was
shown that adding a singlet (real or complex) scalar field, whose potential
mixes with the Higgs field, has important consequences. It turned out that the
RG equations of the combined Higgs-singlet system solves the stabilization
problem faced with a light Higgs field of the order of $125$ Gev avoiding
making the Higgs quartic coupling negative at very high energies. Remarkably,
the form of the Higgs-singlet potential proposed recently agree with the one
we derived before from the spectral action \cite{framework}. The quartic
couplings are determined at unification scale in terms of the gauge and Yukawa
couplings. Running these relations down with the scale, give values consistent
with the present data for the Higgs and top quark mass.

In this note we have analyzed the Higgs-singlet potential resulting from the
spectral action with a cutoff function. We have shown that the quartic Higgs
couplings of the Higgs doublet and singlet get mixed, resulting in shifting
the masses of these two fields. One field, mostly composed of the Higgs get
shifted down, and the one mostly made of sigma get shifted up. We have shown
that it is possible to have a choice of initial conditions, consistent with
the low value of around $125$ Gev for the Higgs mass and $170$ Gev for the top
quark mass.

The lesson we learned from this analysis is that we have to take all the
fields of the noncommutative spectral model seriously, without making
assumptions not backed up by valid analysis, especially because of the almost
uniqueness of the Standard Model in the noncommutative setting. In this
respect this should motivate us to address the remaining questions in the
noncommutative Standard Model. In particular it is important to resolve the
issue of providing a way to make the three gauge couplings meet at some
unification scale. There are other important questions to study in order to
provide a geometric framework for the Yukawa couplings for all fermions, as
well as an explanation for the number of families.

\section{Acknowledgments}A.~H.~C. is supported in part by the National Science
Foundation under Grant No. Phys-0854779.


\begin{thebibliography}{99}                                                                                               %

\bibitem {Why}Ali H. Chamseddine and Alain Connes, \textit{Why the Standard
Model, }J. Geom. Phys. \textbf{58 }(2008) 38-47.

\bibitem {framework}Ali H. Chamseddine and Alain Connes,
\textit{Noncommutative Geometry as a Framework for Unification of all
Fundamental Interactions including Gravity. Part I, }Fortsch. Phys. \textbf{58
}(2010) 553-600.

\bibitem {Elias1}J. Elias-Miro, J. Espinosa, G. Guidice, H. M. Lee and A.
Sturmia, \textit{Stabilization of the Electroweak Vacuum by a Scalar Threshold
effect, }JHEP \textbf{1206 }(2012) 031.

\bibitem {Elias2}G. Degrassi, S. Di Vita, J. Elias-Miro, J. Espinosa, G.
Guidice, G. Isidori and A. Sturmia, \textit{Higgs mass and Vacuum Stability in
the Standard Model at NNLO arXiv:1205.6497.}

\bibitem {Darkon}Chian-Shu Chen and Yong Tang, \textit{Vacuum Stability,
Neutrinos and Dark matter }JHEP \textbf{1204 }(2012) 019.

\bibitem {lebedev}Oleg Lebedev \textit{On Stability of the Higgs Potential and
the Higgs Portal, JHEP, }arXiv:1203.0156.
\bibitem{Bezr} 
  F.~Bezrukov, M.~Y.~.Kalmykov, B.~A.~Kniehl and M.~Shaposhnikov,
  arXiv:1205.2893 [hep-ph].

\bibitem{RGEsing}
 M.~Gonderinger, Y.~Li, H.~Patel and M.~J.~Ramsey-Musolf,
  JHEP {1001} (2010) 053
  [hep-ph/0910.3167];
  O.~Lebedev and H.~M.~Lee,
  Eur.\ Phys.\ J.\ C {71} (2011) 1821
  [hep-ph/1105.2284];
  M.~Kadastik, K.~Kannike, A.~Racioppi and M.~Raidal,
  [hep-ph/1112.3647];
  M.~Gonderinger, H.~Lim and M.~J.~Ramsey-Musolf,
  [hep-ph/1202.1316];
 C.~-S.~Chen and Y.~Tang
  [hep-ph/1202.5717].



\bibitem {mc2}A.~Chamseddine, A.~Connes, M.~Marcolli, \emph{Gravity and the
standard model with neutrino mixing}, Adv. Theor. Math. \textbf{11} (2007) 991-1090.



\bibitem {ali}Ali H. Chamseddine, \textit{Noncommutative Geometry as the key
to unlock secrets of space-time, }Published in Quanta of Math, editors E.
Blanchard et al, Clay Mathematics Institute/AMS publication, 2010,
arXiv:0901.0577.

\bibitem {Zoller}K. Chetyrkin and M. Zoller, \textit{Three loop beta functions
for the top Yukawa and the Higgs self interactions in the Standard Model},
arXiv:1205.2892.

\bibitem {scale}Ali H. Chamseddine and Alain Connes, \textit{Scale invariance
in the spectral action, }J. Math. Phys. \textbf{47 }(2006) 063504.
\bibitem{HiggsBoson}
 F.~Englert and R.~Brout,
Phys.\ Rev.\ Lett.\ \ {\bf 13}, 321  (1964);
  P.~W.~Higgs,
Phys.\ Lett.\ \ {\bf 12}, 132  (1964).
Phys.\ Rev.\ Lett.\ \ {\bf 13}, 508  (1964);
  G.~S.~Guralnik, C.~R.~Hagen and T.~W.~B.~Kibble,
Phys.\ Rev.\ Lett.\ \ {\bf 13}, 585  (1964).
\end{thebibliography}
\end{document}